\documentclass[reprint,twocolumn,notitlepage,preprintnumbers,nofootinbib,amsmath,amssymb,aps,prd,floatfix,superscriptaddress,longbibliography,svgnames]{revtex4-1}

\usepackage{amsmath}
\usepackage{amsfonts}
\usepackage{amssymb}
\usepackage{graphicx}
\usepackage{bm}
\usepackage{subfigure}
\usepackage{url}
\usepackage[hyperindex]{hyperref}
\usepackage{color}
\usepackage[ddmmyy,24hr]{datetime}
\usepackage{bigdelim}
\usepackage{booktabs}
\usepackage{dcolumn}
\usepackage{multirow}
\usepackage{subfigure}
\usepackage{physics}
\usepackage{cancel}
\usepackage{stackrel}
\usepackage{paralist}
\usepackage{xspace}
\usepackage{slashed}
\usepackage{cancel}
\usepackage{todonotes}
\usepackage{enumerate}
\usepackage{float}
\usepackage{fullpage}
\usepackage{ulem}
\usepackage{wasysym}
\usepackage{comment}
\usepackage{bbold}
\usepackage{hyperref}
\newcommand{\cenns}{CE$\nu$NS }
\usepackage{xcolor}
\usepackage{orcidlink}

\usepackage{stackengine}

\usepackage{feynmp-auto}
\newcommand{\nua}[1]{\ensuremath{\rlap{\kern-2.5pt\ensuremath{\overset{\scriptscriptstyle(-)}{\phantom{\nu}}}}{\ensuremath{{\nu}_{#1}}}}}

\usepackage{enumitem}

\begin{document}

\title{Phenomenological implications of the high-precision COHERENT germanium CE$\nu$NS data}

\author{M. Atzori Corona \orcidlink{0000-0001-5092-3602}}
\email{mcorona@roma2.infn.it}
\affiliation{Istituto Nazionale di Fisica Nucleare (INFN), Sezione di Roma Tor Vergata, Via della Ricerca Scientifica, I-00133 Rome, Italy}

\author{M. Cadeddu \orcidlink{0000-0002-3974-1995}}
\email{matteo.cadeddu@ca.infn.it}
\affiliation{Istituto Nazionale di Fisica Nucleare (INFN), Sezione di Cagliari,
	Complesso Universitario di Monserrato - S.P. per Sestu Km 0.700,
	09042 Monserrato (Cagliari), Italy}

\author{N. Cargioli \orcidlink{0000-0002-6515-5850}}
\email{nicola.cargioli@ca.infn.it}
\affiliation{Istituto Nazionale di Fisica Nucleare (INFN), Sezione di Cagliari,
	Complesso Universitario di Monserrato - S.P. per Sestu Km 0.700,
	09042 Monserrato (Cagliari), Italy}

\author{R. Cerulli \orcidlink{https://orcid.org/0000-0003-2051-3471}}
\email{riccardo.cerulli@roma2.infn.it}
\affiliation{Istituto Nazionale di Fisica Nucleare (INFN), Sezione di Roma Tor Vergata, Via della Ricerca Scientifica, I-00133 Rome, Italy}
\affiliation{Dipartimento di Fisica, Università di Roma ‘Tor Vergata’, I-00133 Rome, Italy}

\author{\mbox{G. Co' \orcidlink{0000-0002-9613-5211}}}
\email{giampaolo.co@le.infn.it}
\affiliation{Dipartimento di Matematica e Fisica ``E. De Giorgi''
Università del Salento.\\
Istituto Nazionale di Fisica Nucleare (INFN), Sezione di Lecce\\
via Arnesano, 73100 Lecce, Italy
}

\author{F. Dordei \orcidlink{0000-0002-2571-5067}}
\email{francesca.dordei@cern.ch}
\affiliation{Istituto Nazionale di Fisica Nucleare (INFN), Sezione di Cagliari,
	Complesso Universitario di Monserrato - S.P. per Sestu Km 0.700,
	09042 Monserrato (Cagliari), Italy}

\author{C. Giunti \orcidlink{0000-0003-2281-4788}}
\email{carlo.giunti@to.infn.it}
\affiliation{Istituto Nazionale di Fisica Nucleare (INFN), Sezione di Torino, Via P. Giuria 1, I--10125 Torino, Italy}

\author{R. Pavarani \orcidlink{0009-0004-9534-542X}}
\email{riccardo.pavarani@ca.infn.it}
\affiliation{Istituto Nazionale di Fisica Nucleare (INFN), Sezione di Cagliari,
	Complesso Universitario di Monserrato - S.P. per Sestu Km 0.700,
	09042 Monserrato (Cagliari), Italy}
\affiliation{Università degli Studi di Padova, Dipartimento di Fisica e Astronomia “Galileo Galilei”, Via Francesco Marzolo 8, 35131 Padova, Italy}

\date{\dayofweekname{\day}{\month}{\year} \ddmmyydate\today, \currenttime}

\begin{abstract}
This work presents the first comprehensive phenomenological analysis of the newly released Coherent Elastic Neutrino-Nucleus Scattering (CE$\nu$NS) data on germanium, measured by the COHERENT collaboration at the Spallation Neutron Source. Leveraging the unprecedented precision of this dataset, we provide state-of-the-art determinations of key Standard Model and nuclear physics parameters. Specifically, we extract updated constraints on the weak mixing angle, the neutrino charge radii, and we perform a detailed extraction of the neutron root-mean-square radius of germanium nuclei. We also investigate the impact of quenching factor uncertainties by exploring an extended Lindhard framework, and assess their effect on the extraction of nuclear parameters.
Additionally, we use these results to evaluate scenarios beyond the Standard Model, placing robust bounds on neutrino non-standard interactions. To maximize the statistical power and robustness of our findings, whenever possible, we perform a global combined analysis incorporating previous COHERENT measurements along with reactor antineutrino data from the CONUS+, TEXONO, and $\nu$GeN experiments as well as dark-matter experiments.
\end{abstract}

\maketitle  

\section{Introduction}

The transition from a statistics-dominated to a systematics-dominated measurement marks an important threshold in any experimental program. 
It signals that the dataset has matured to the point where further progress requires not only more data, but deeper control over 
the underlying physics. The COHERENT collaboration has recently crossed this threshold in the coherent elastic neutrino-nucleus scattering 
(CE$\nu$NS) program on germanium~\cite{COHERENT:2026yje}. 
With roughly three times the neutrino exposure of their previous result~\cite{COHERENT:2025vuz}, an active mass of \mbox{$(8.53\pm0.08)\;\rm kg$}, a lowered analysis threshold of 
$0.5~\rm{keV}$, made possible by improved pulse onset reconstruction, and a substantial suppression of backgrounds through pulse shape discrimination, the new Ge-Mini measurement achieves a total uncertainty of approximately 10\%, with the neutrino flux normalization now constituting the dominant source of systematic uncertainty. This is the most precise CE$\nu$NS measurement to date.

From a phenomenological standpoint, this development calls for a thorough reanalysis. The first COHERENT germanium dataset~\cite{COHERENT:2025vuz}, which has been 
studied in Refs.~\cite{AtzoriCorona:2025xgj,Li:2025pfw,Liao:2024qoe}, was limited by statistics at the 30\% level, and exhibited a mild deficit with respect to the Standard Model (SM) prediction whose interpretation remained ambiguous. The new measurement resolves this tension: the observed signal yield is fully consistent with the 
SM within $1\sigma$, both in the new dataset alone and in the combination with the earlier result. 

CE$\nu$NS~\cite{Freedman:1973yd} is a low-energy weak neutral-current process whose cross section scales with the square of the nuclear weak 
charge, making it simultaneously sensitive to the electroweak sector of the SM, to the internal structure of nuclei, and to new physics modifying neutrino couplings~\cite{DeRomeri:2026prc,DeRomeri:2024hvc,Coloma:2022avw,Demirci:2023tui,Demirci:2024vzk,DeRomeri:2022twg,DeRomeri:2026kgg}. At the Spallation Neutron Source (SNS), the well-determined timing structure of the neutrino flux further separates the prompt $\nu_\mu$ component from the delayed $\bar{\nu}_\mu$ and $\nu_e$ contributions, enabling independent constraints on flavor-dependent interactions. All of these handles become sharper as the statistical uncertainty shrinks, motivating a new round of precision extractions of the weak mixing angle $\sin^2\vartheta_W$, the neutron root-mean-square (rms) radius of germanium, neutrino charge radii, and bounds on non-standard neutrino interactions (NSI).

The phenomenological reach of the new dataset is further amplified by the availability of complementary CE$\nu$NS measurements on the same 
nuclear target from reactor antineutrino experiments, such as CONUS+~\cite{Ackermann:2025obx}. At reactor sites, the neutrino energies are limited to a few MeV, keeping the momentum 
transfer small enough that nuclear form factor effects are negligible, thereby providing a cleaner window onto fundamental SM parameters and 
neutrino properties. This complementarity between accelerator and reactor sources on a common germanium target, first exploited in a previous analysis~\cite{AtzoriCorona:2025xgj}, is now considerably more powerful given the improved precision of the SNS measurement. Moreover, we incorporate data from COHERENT CsI~\cite{Akimov:2021dab} and 
Ar~\cite{COHERENT:2020iec, COHERENT:2020ybo},  TEXONO~\cite{TEXONO:2024vfk}, and $\nu$GeN~\cite{nGeN:2025hsd} as well as dark-matter experiments~\cite{XENON:2024ijk,PandaX:2024muv,LZ:2025igz} into global combined analyses wherever the combination adds discriminating power.
These results demonstrate once again the power of CE$\nu$NS as a fundamental probe of the SM, which is expected to further improve in the near future thanks to new detectors and facilities under development~\cite{Chatterjee:2022mmu,Collar:2025sle,Chattaraj:2025rtj,NUCLEUS:2026utx,AtzoriCorona:2025ibl,NUCLEUS:2026pnv}\footnote{During the completion of this work, an independent study presenting a similar phenomenological analysis of the new COHERENT germanium data appeared in Ref.~\cite{DeRomeri:2026dac}.}.

\section{CE$\nu$NS cross section and nuclear inputs}
\label{sec:theory}

The differential cross section for CE$\nu$NS of a neutrino 
$\nu_\ell$ ($\ell = e, \mu, \tau$) off a nucleus $\mathcal{N}$ 
with $Z$ protons and $N$ neutrons, as a function of the nuclear 
recoil energy $T_\mathrm{nr}$, reads~\cite{Cadeddu:2023tkp}
\begin{equation}
    \frac{d\sigma_{\nu_{\ell}\text{-}\mathcal{N}}}{d T_\mathrm{nr}} = 
    \frac{G_{\text{F}}^2 M}{\pi} 
    \left( 1 - \frac{M T_\mathrm{nr}}{2 E^2} \right)
    \left( Q^{V}_{\ell, \mathrm{SM}} \right)^2,
    \label{eq:cexsec}
\end{equation}
where $G_{\text{F}}$ is the Fermi constant, $E$ the neutrino energy, 
and $M$ the nuclear mass.\footnote{We use natural units $\hbar = c = 1$ 
throughout. For the isotopic composition of germanium we use the values 
from Ref.~\cite{BerglundWieser+2011+397+410}. Here, we neglect the very small axial-vector contribution to the interaction~\cite{AbdelKhaleq:2024hir,AristizabalSierra:2026rlo}.} The weak nuclear charge 
is
\begin{equation}
    Q^{V}_{\ell, \mathrm{SM}} = g_{V}^{p}(\nu_\ell)\, Z F_Z(|\vec{q}|^2) 
    + g_{V}^{n}\, N F_N(|\vec{q}|^2),
    \label{eq:weakcharge}
\end{equation}
where $g_{V}^{p}(\nu_\ell)$ and $g_{V}^{n}$ are the weak neutral-current 
vector couplings of protons and neutrons, and $F_Z$, $F_N$ are the 
corresponding nuclear form factors evaluated at momentum transfer 
$|\vec{q}| \simeq \sqrt{2MT_{\rm{nr}}}$. 
Including radiative corrections in the $\overline{\mathrm{MS}}$ 
scheme~\cite{AtzoriCorona:2023ktl, Erler:2013xha, ParticleDataGroup:2024cfk}, 
the SM values of these couplings are~\cite{AtzoriCorona:2025ygn}
\begin{align}
g_{V}^{p}(\nu_{e}) &= 0.0379, \quad g_{V}^{p}(\nu_{\mu}) = 0.0297, \\
g_{V}^{p}(\nu_{\tau}) &= 0.0253, \quad g_{V}^{n} = -0.5117,
\end{align}
obtained fixing $\sin^2\vartheta_W$ to its SM value at low momentum transfer, \mbox{$\sin^2\vartheta_W(Q \to 0) = 0.23873(5)$~\cite{ParticleDataGroup:2024cfk}}.
The flavor dependence of $g_{V}^{p}(\nu_\ell)$ originates from the 
neutrino charge radii, which represent the only non-vanishing 
electromagnetic properties of neutrinos in the 
SM~\cite{Giunti:2024gec}.

The nuclear form factors in Eq.~(\ref{eq:weakcharge}) encode the 
spatial distribution of protons and neutrons inside the nucleus via 
the Fourier transform of the corresponding nucleon density 
$\rho_{Z(N)}$. Their departure from unity grows with increasing 
momentum transfer and produces a progressive suppression of full 
coherence~\cite{Cadeddu:2017etk}, that is central to the 
interpretation of COHERENT data.
To describe the form factors, the Helm parameterisation~\cite{Helm:1956zz} could be adopted for both 
$F_Z$ and $F_N$, which has been shown to be practically equivalent 
to the symmetrised Fermi~\cite{Piekarewicz:2016vbn} and 
Klein-Nystrand~\cite{Klein:1999qj} parameterisations. The Helm 
form factor depends on the proton and neutron rms radii of the 
nucleus, and we fix the surface thickness to $s=0.9\;\rm fm$~\cite{Friedrich:1982esq}. Proton rms radii are derived from charge radii obtained via muonic atom spectroscopy and electron scattering~\cite{Fricke:1995zz, Fricke2004, Angeli:2013epw,Cadeddu:2020lky}, whose average value for the isotopic composition of germanium is equal to $R_p(\rm Ge)=4.078(1)\;\rm fm$. For the neutron rms radii, which lack precise 
direct measurements for the germanium isotopes, one could use the values obtained from the recent nuclear shell model (NSM) estimate of the corresponding neutron skins $\Delta R_{\rm np}$ (i.e. the differences between the neutron and the proton rms radii) in Ref.~\cite{Hoferichter:2020osn}.
Here, however, as our baseline, we adopt the 
predictions from Hartree-Fock plus Bardeen-Cooper-Schrieffer 
(HF+BCS) calculations~\cite{Co:2021ijy} using the D1S Gogny 
effective interaction~\cite{Berger:1991zza}. To quantify the sensitivity to nuclear modelling, we assess the spread between 
this baseline and two alternative predictions: the aforementioned NSM estimate, which 
tends to yield larger neutron radii, and the HF+BCS calculation 
with the D1M Gogny interaction~\cite{Chabanat:1997un}, which 
gives the smallest values. The proton and neutron rms radii 
for all models and germanium isotopes are summarised in 
Tab. I of Ref.~\cite{AtzoriCorona:2025xgj}.

\section{Analysis of CE$\nu$NS data}
\label{sec:analysis}

\subsection*{Existing datasets}

For the reactor-based experiments, we analyse CONUS+~\cite{Ackermann:2025obx} 
and TEXONO~\cite{TEXONO:2024vfk} data following 
Ref.~\cite{AtzoriCorona:2025ygn}, and $\nu$GeN 
data~\cite{nGeN:2025hsd} following Ref.~\cite{AtzoriCorona:2025xgj}. 
We refer to their combination hereafter as \texttt{Reactors}. 
The COHERENT CsI~\cite{Akimov:2021dab} and 
Ar~\cite{COHERENT:2020iec, COHERENT:2020ybo} datasets are 
analysed according to Refs.~\cite{AtzoriCorona:2023ktl, 
AtzoriCorona:2022moj}, and the first 2025 COHERENT germanium 
dataset~\cite{COHERENT:2025vuz} according to 
Ref.~\cite{AtzoriCorona:2025xgj}.

\subsection*{COHERENT germanium 2026}\label{sec:IntroCohGe}

The neutrino flux at the SNS consists of a prompt $\nu_\mu$ 
component and a delayed $\nu_e$ and $\bar{\nu}_\mu$ component 
with well-determined energy spectra~\cite{COHERENT:2026yje}. 
The flux is characterised by $N_{\mathrm{POT}} = 4.68 \times 10^{22}$ 
protons on target, a yield of $(0.37 \pm 0.04)$ neutrinos per proton, 
and a source-to-detector baseline of $L = 19.2$~m~\cite{COHERENT:2026yje}.

In each nuclear-recoil energy-bin $i$, 
the expected \cenns event number $N^\mathrm{CE \nu NS}_{i}$ on a germanium target is given by
\begin{align}\label{N_cevns}\nonumber
N_{i}^{\mathrm{CE}\nu\mathrm{NS}}
&=
N_T
\int_{T_{\mathrm{nr}}^{i}}^{T_{\mathrm{nr}}^{i+1}}
\hspace{-0.3cm}
d T_{\mathrm{nr}}\,
\int_{0}^{T^{\prime\text{max}}_{\text{nr}}}
\hspace{-0.3cm}
dT'_{\text{nr}}
\,
R(T_{\text{nr}},T'_{\text{nr}})\\
&\times\! \int_{E_{\text{min}}(T'_{\text{nr}})}^{E_{\text{max}}}
\hspace{-0.3cm}
d E
\sum_{\nu_{\ell}=\nu_{e}, \nu_{\mu}, \bar{\nu}_{\mu}}
\frac{d N_{\nu_{\ell}}}{d E}
\frac{d \sigma_{\nu_{\ell}-\mathcal{N}}}{d T'_{\mathrm{nr}}},
\end{align}
with the sum running over $\nu_\ell \in \{\nu_e, \nu_\mu, \bar\nu_\mu\}$. 
Here $dN_{\nu_\ell}/dE$ is the neutrino flux~\cite{COHERENT:2021yvp}, $R(T_\mathrm{nr}, 
T'_\mathrm{nr})$ is the energy resolution function~\cite{COHERENT:2025vuz}, 
$T^{\prime\text{max}}_\mathrm{nr} \simeq 2E_\mathrm{max}^2/M$, 
$E_\mathrm{max} = m_\mu/2 \simeq 52.8$~MeV, 
$E_\mathrm{min}(T'_\mathrm{nr}) \simeq \sqrt{MT'_\mathrm{nr}/2}$, 
and $N_T$ is the total number of target atoms in the detector volume.

The relation between the observed ionization energy $T_e$ and the 
true nuclear recoil energy $T_\mathrm{nr}$ is given by 
$T_e = f_Q(T_\mathrm{nr})\, T_\mathrm{nr}$, where $f_Q$ is the 
quenching factor, whose energy dependence has important consequences 
for the extraction of SM and beyond-SM parameters~\cite{Colaresi:2022obx, 
AtzoriCorona:2023ais, Li:2025pfw}. 

In this work, unless otherwise specified, we adopt the Lindhard model with $k = 0.157$~\cite{Lindhard_theo}, while we will explore possible deviations from this standard paradigm in Sec.~\ref{sec:QFModels}.
The timing structure of the data is incorporated by extracting 
the arrival-time distributions of the individual neutrino 
components from Ref.~\cite{GermanioWineAndCheese} and distributing 
the predicted CE$\nu$NS events into time bins of $2~\mu$s, 
yielding $N^{\mathrm{CE}\nu\mathrm{NS}}_j$ where $j$ labels the 
time bins.

\begin{figure}[t]
    \includegraphics[width=\linewidth]{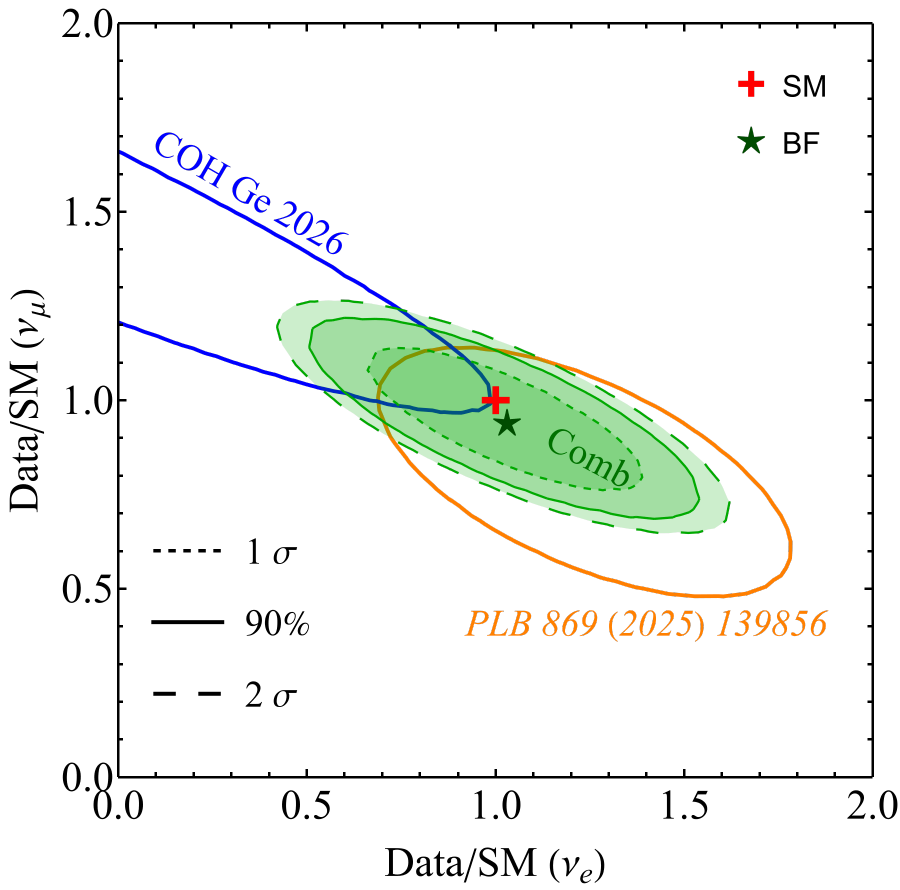}
    \caption{Agreement between CE$\nu$NS data and SM predictions for COHERENT germanium 2026 data (blue contour), the combination~\cite{AtzoriCorona:2025xgj} of CsI+Ar, \texttt{Reactors}, and COHERENT germanium 2025 data (orange 
    contour) and the full combination (green regions), displayed 
    separately for electron and muon neutrinos. Contours are at 1$\sigma$ (dotted), 
    90\% (solid) and 2$\sigma$ (dashed) CL. The dark green star marks the best fit, and the red cross 
    indicates the SM expectation.}
    \label{fig:Norm2D}
\end{figure}
Since the event-by-event energy-time distribution is not publicly available, a full two-dimensional likelihood analysis cannot be performed. Consequently, this analysis relies on independent one-dimensional fits, which inherently neglect the correlations between the energy and time datasets. However, we have verified that the impact of this approximation is negligible, as the systematic uncertainty dominates over the statistical one. Therefore, the two datasets can be safely combined to fully exploit the complementary information encoded in the CE$\nu$NS signal, namely, its recoil-energy spectrum and timing structure, without affecting the extraction of the physical parameters of interest.
The analysis is thus performed via a simultaneous least-squares 
fit to the beam-on (ON) and beam-off (OFF) data in both energy (E) and 
\mbox{time (T)}, treating the four one-dimensional distributions independently:
\begin{align}\label{eq:chi2tot}
\chi^2 &= \chi^2_{\text{E,ON}}(\eta,\beta) 
        + \chi^2_{\text{E,OFF}}(\beta) \nonumber\\
       &+ \chi^2_{\text{T,ON}}(\eta,\beta) 
        + \chi^2_{\text{T,OFF}}(\beta)
        + \left(\frac{\eta - 1}{\sigma_\eta}\right)^2,
\end{align}
where each term is a Poissonian least-squares function. The energy ON and OFF contributions are given by
\begin{align}\nonumber\label{eq:chi2poiss}
\chi^2_{\text{E,ON}} &= 2 \sum_{\substack{i}}\bigg( 
\eta N^{\text{CE}\nu\text{NS}}_{i} 
+ \beta N^{\text{SSB}}_{i} 
- N^{\text{exp}}_{i} \\
&\quad + N_{i}^{\text{exp}} 
\mathrm{ln} \left[ \frac{N_{i}^{\text{exp}}}
{\eta N^{\text{CE}\nu\text{NS}}_{i} 
+ \beta N^{\text{SSB}}_{i} } \right] \bigg),
\end{align}
\begin{equation}\label{eq:chi2poiss}
\chi^2_{\text{E,OFF}} = 2 \sum_{\substack{i}}\bigg( 
\beta N^{\text{SSB}}_{i} 
- N^{\text{exp}}_{i} + N_{i}^{\text{exp}} 
\mathrm{ln} \left[ \frac{N_{i}^{\text{exp}}}
{ \beta N^{\text{SSB}}_{i} } \right] \bigg),
\end{equation}
with $N_i^{\rm exp}$ the observed counts of the corresponding data-set as taken from Ref.~\cite{COHERENT:2026yje}, 
and $N_i^{\rm SSB}$ the steady-state background prediction in \mbox{bin $i$}. Analogous expressions are adopted for the timing  $\rm ON$ and  $\rm OFF$ datasets, replacing the energy-binned event distributions with the corresponding time-binned distributions.
The nuisance parameter $\eta$ rescales the signal and is shared 
across all the $\rm ON$ datasets, while $\beta$ accounts for the background normalization and is shared across all four datasets. The beam-off datasets are used to constrain the nuisance parameter $\beta$, thereby fixing the steady-state background contribution entering the beam-on likelihood. The signal systematic uncertainty is $\sigma_\eta = 10.3\%$, 
dominated by the neutrino flux normalization (10\%), with subdominant 
contributions from the detector baseline (0.5\%), energy calibration 
(0.8\%), active mass (2.0\%), form factors (0.8\%) and quenching 
factor (0.7\%)~\cite{COHERENT:2026yje}.

Our fit to the total CE$\nu$NS normalization yields $127^{+10}_{-9}$ events, 
in good agreement with the COHERENT result of $124^{+14}_{-12}$~\cite{COHERENT:2026yje} and with our theoretical prediction of $129$ events under the D1S model. 
It should be noted that the quoted uncertainty is purely statistical, excluding the 10.3\% systematic uncertainty on the overall CE$\nu$NS normalization. While performing independent 1D fits leads to a slight underestimation of the statistical uncertainty compared to the official COHERENT analysis, this effect becomes completely negligible once the dominant systematic uncertainty is incorporated in the extraction of the physical parameters.

To probe potential flavor-dependent effects and exploit the 
discriminating power of the timing information, we split the 
signal normalization \mbox{$\eta = \mathrm{Data/SM}$} into independent 
electron- and muon-neutrino components following the strategy 
introduced in Ref.~\cite{AtzoriCorona:2025ygn}. This decomposition 
also enables a direct comparison with \texttt{Reactor} data, which are 
sensitive exclusively to the $\nu_e$ flavor. Putting together 
COHERENT Ge~2026 with COHERENT CsI, Ar and \texttt{Reactors}, the results of this combined analysis are shown in Fig.~\ref{fig:Norm2D} 
and yield
\begin{align}\nonumber
\eta\,(\nu_e)\! &=\! 
1.03^{+0.24}_{-0.25}\,(1\sigma),\;
^{+0.39}_{-0.40}\,(90\%),\;
^{+0.48}_{-0.49}\,(2\sigma),\\
\eta\,(\nu_\mu)\! &=\!
0.94^{+0.14}_{-0.12}\,(1\sigma),\;
^{+0.22}_{-0.20}\,(90\%),\;
^{+0.26}_{-0.24}\,(2\sigma).
\end{align}
Both values are consistent with the SM prediction within $1\sigma$, resolving the mild tension observed in the earlier COHERENT germanium data~\cite{COHERENT:2025vuz, AtzoriCorona:2025xgj}. 
A general agreement among all CE$\nu$NS datasets and probes is 
observed across the full combination.
A closer inspection of the COHERENT Ge~2026 contour alone reveals that, despite the good agreement of the total event yield discussed above, the data are only marginally consistent with the SM prediction when the two flavor components are treated independently, lying approximately at the boundary of the 90\% CL 
region. This feature originates from a mild excess of events in the prompt timing window, which is populated predominantly by $\nu_\mu$ neutrinos, driving a preference for an enhanced muon-neutrino cross section. This effect is absorbed into the combined analysis and is consistent with a statistical fluctuation, but it is worth noting that it will persist in all analyses that allow for flavor-dependent signal normalizations.

\section{Implications for Electroweak and Nuclear Physics}
\label{sec:SM}

\subsection{Neutron rms radius of germanium}
\label{sec:nuclear}

We extract the neutron rms radius of germanium $R_n(\mathrm{Ge})$, averaged over the isotopic composition, 
by treating it as a free parameter in the fit while 
fixing $\sin^2\vartheta_W$ to its SM value at low momentum transfer. In this fit, the $0.8\%$ systematic uncertainty associated with the nuclear form factors is removed from the overall systematic error budget of the CE$\nu$NS prediction. The proton rms radii are held fixed to the values 
derived from the measured charge radii, as described in Sec.~\ref{sec:theory}. 
\begin{figure}[t]
    \includegraphics[width=\linewidth]{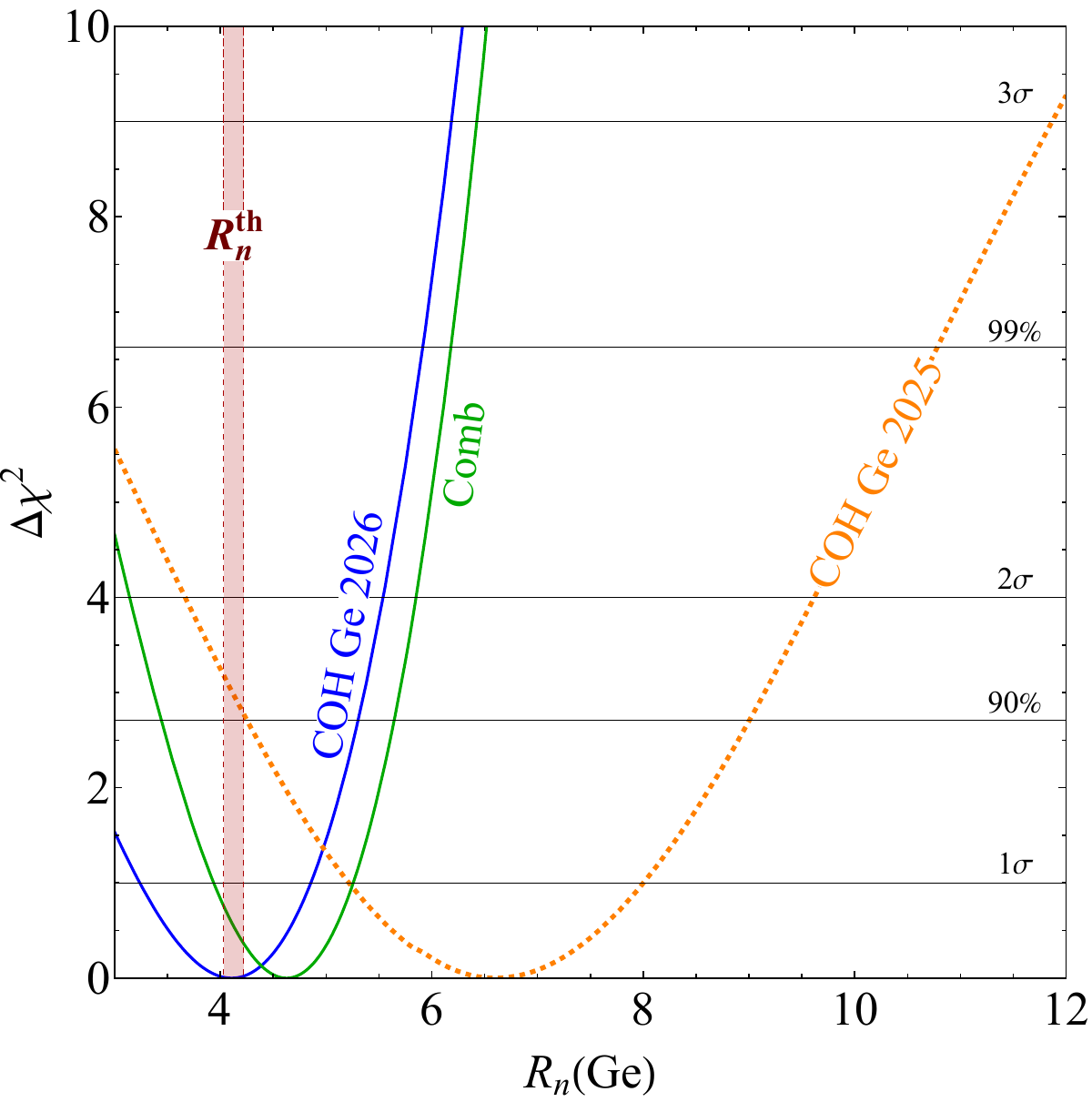}
    \caption{$\Delta\chi^2$ profile as a function of $R_n(\mathrm{Ge})$ for COHERENT Ge~2026 (blue solid line) and, for comparison, for the first COHERENT Ge~2025 (orange dotted line)~\cite{AtzoriCorona:2025xgj}. The combined analysis is shown in green, while the red vertical band represents the range of the average germanium neutron radius predicted by nuclear models, spanning from the D1M interaction~\cite{Chabanat:1997un} to the NSM model~\cite{Hoferichter:2020osn}.}
    \label{fig:Rn_profile}
\end{figure}

The resulting $\Delta\chi^2$ profile as a function of $R_n(\mathrm{Ge})$ 
is shown in Fig.~\ref{fig:Rn_profile} for COHERENT Ge~2026 
and, for comparison, the first COHERENT 2025 germanium~\cite{COHERENT:2025vuz,AtzoriCorona:2025xgj}
datasets.
The theoretical $R_n^{\mathrm{th}}$ prediction is indicated by the red vertical band, following the prescription described in 
Sec.~\ref{sec:theory}.

The improvement brought 
by the new data is immediately apparent: the 2026 profile is 
substantially narrower, with the $1\sigma$ uncertainty reduced 
by roughly a factor of two with respect to the earlier result. 
The best-fit value and associated $1\sigma$ uncertainties read
\begin{equation}
 \textrm{COH Ge 2026:} \quad   R_n(\mathrm{Ge}) = 4.11^{+0.75}_{-0.87}\;\mathrm{fm}\,,
\end{equation}
consistent with the D1S theoretical prediction at the $0.1\sigma$ level. We also show 
the combined result obtained by summing the profile likelihoods 
of both germanium datasets, which further tightens the constraint 
to 
\begin{equation}
\label{eq:Rn_comb}
 \textrm{Comb:} \,  
R_n(\mathrm{Ge}) = 4.63^{+0.63}_{-0.69}~\mathrm{fm}\,,
\end{equation}
lying about $0.8\sigma$ from the theoretical prediction, indicated as the spread between the average NSM estimate ($R_n = 4.22\,\mathrm{fm}$) and the
HF+BCS calculation with the D1M Gogny interaction ($R_n = 4.03\,\mathrm{fm}$). This result can be directly translated into a measurement of the 
average neutron skin of germanium
\begin{equation}
\Delta R_{\rm np}(\mathrm{Ge}) = 0.54^{+0.63}_{-0.69}~\mathrm{fm}.
\end{equation}
While the uncertainty remains large, the central value exceeds the nuclear model predictions of $\Delta R_{\rm np}^{\rm th} \sim \left[0.07-0.14\right]$~fm~\cite{AtzoriCorona:2025xgj}. 
This trend is not unique to germanium: similarly large neutron skin values have been extracted from CE$\nu$NS measurements on CsI and Ar targets~\cite{AtzoriCorona:2023ktl,Cadeddu:2020lky}, and are also observed in electroweak probes of heavy nuclei such as 
PREX-II~\cite{PREX:2021umo}. Whether this pattern reflects a genuine systematic effect in the CE$\nu$NS analyses, a common bias in nuclear form factor modelling, or simply the limited statistical power of current datasets remains an open question that future high-precision measurements will be able 
to address.

\subsection{Weak mixing angle}
\label{sec:sin2w}

\begin{figure}[th]
\includegraphics[width=\linewidth]{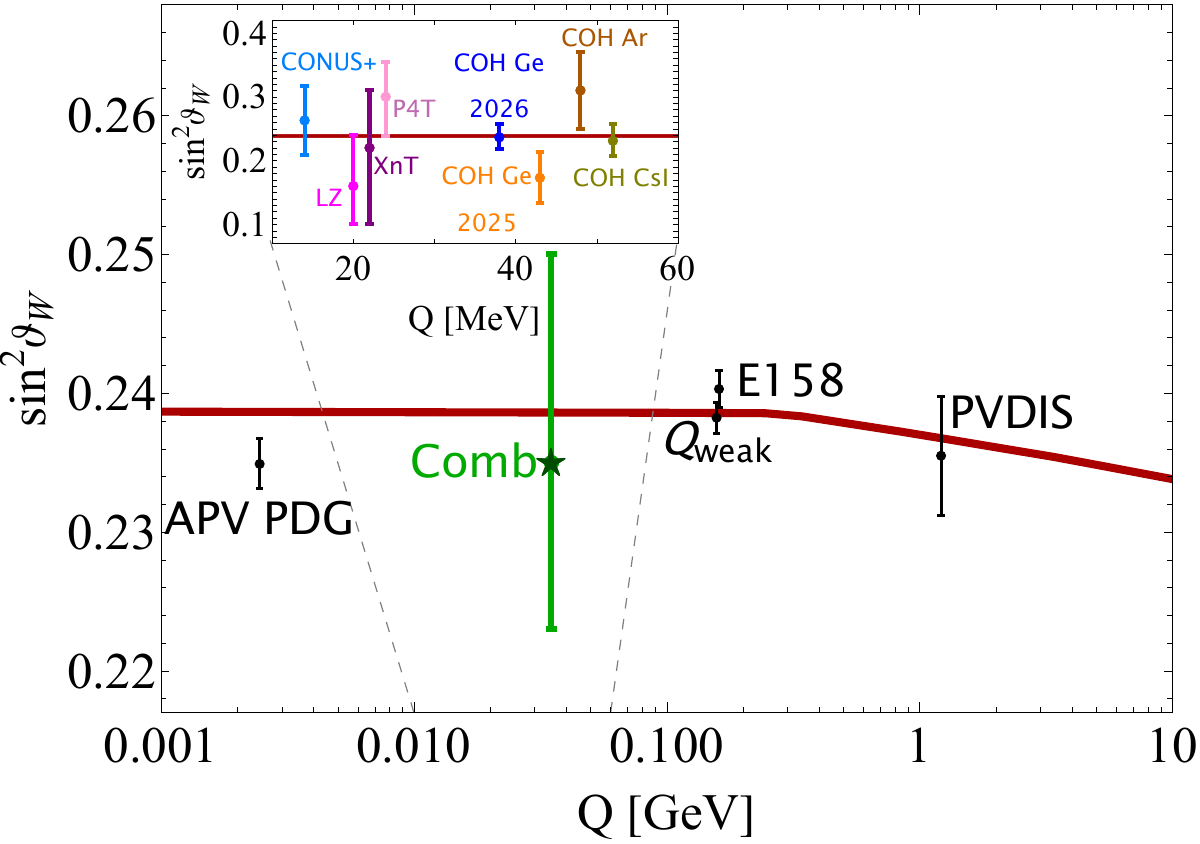}
    \caption{Variation of the value of $\sin^2\vartheta_W$ with $Q$ in the low-energy range. The SM prediction is shown as the dark red solid curve, together with experimental determinations in black from Møller scattering (E158)~\cite{Anthony:2005pm}, deep inelastic scattering of polarized electrons on deuterons (PVDIS)~\cite{Wang:2014bba}, and the result from the proton weak charge (Qweak)~\cite{Androic:2018kni} and atomic parity violation~\cite{ParticleDataGroup:2024cfk}. The result derived in this paper for COHERENT Germanium 2026 data is shown in blue in the inset, together with the COHERENT CE$\nu$NS measurements from CsI~\cite{AtzoriCorona:2023ktl}, Ar~\cite{Cadeddu:2020lky}, and Germanium 2025~\cite{AtzoriCorona:2025xgj} datasets, CONUS+~\cite{AtzoriCorona:2025ygn}, PandaX~\cite{AtzoriCorona:2025gyz},  XENONnT~\cite{AtzoriCorona:2025gyz}, and LZ~\cite{LZ:2025igz}. The combined analysis of all CE$\nu$NS data is reported in green.}
    \label{fig:sin2w_Q}
\end{figure}

To place our results in the broader context of 
electroweak precision measurements, we also 
perform a one-dimensional extraction of 
$\sin^2\vartheta_W$ with $R_n(\mathrm{Ge})$ fixed 
to the D1S prediction, corresponding to an average radius of $R_n (\mathrm{Ge}) = 4.09\,\mathrm{fm}$~\cite{AtzoriCorona:2025xgj}. The outcome is displayed 
in Fig.~\ref{fig:sin2w_Q} together with the SM running prediction as a function 
of the characteristic momentum transfer $Q$. 
The new COHERENT germanium determination at $Q \sim 40$~MeV
\begin{equation}
 \textrm{COH Ge 2026:}\quad 
\sin^2\vartheta_W = 0.237^{+0.021}_{-0.019}\,,
\end{equation}
represents the 
most precise single-experiment extraction of the weak 
mixing angle from CE$\nu$NS data  
to date. The improvement over the previous germanium 
result~\cite{COHERENT:2025vuz} reflects directly the 
reduction of the statistical uncertainty from 30\% to 
below 10\% in the new dataset. 
We perform a combined analysis including COHERENT CsI~\cite{Akimov:2021dab}, 
Ar~\cite{COHERENT:2020iec, COHERENT:2020ybo}, and the previous COHERENT 2025 germanium result~\cite{COHERENT:2025vuz}, together with reactor CE$\nu$NS data from CONUS+~\cite{Ackermann:2025obx}, TEXONO~\cite{TEXONO:2024vfk}, and $\nu$GeN~\cite{nGeN:2025hsd} as well as CE$\nu$NS results from XENONnT~\cite{XENON:2024ijk} and PandaX~\cite{PandaX:2024muv}, analyzed as detailed in Ref.~\cite{AtzoriCorona:2025gyz}. For the LZ~\cite{LZ:2025igz} dataset, we adopt a conservative approach by using a Gaussian prior on $\sin^2\vartheta_W$, centered on the collaboration best-fit value and with a standard deviation $\sigma = 0.08$~\cite{LZ:2025igz}.
This global analysis yields 
\begin{equation}
 \textrm{Comb:}\quad 
\sin^2\vartheta_W = 0.235^{+0.015}_{-0.012}\,,
\end{equation}
representing the current state of the art in low-energy determinations of the weak mixing angle based on CE$\nu$NS probes. It is shown in Fig.~\ref{fig:sin2w_Q} alongside determinations from other 
low-energy probes including APV~\cite{ParticleDataGroup:2024cfk}, E158~\cite{Anthony:2005pm}, 
$Q_\mathrm{weak}$~\cite{Androic:2018kni}, and PVDIS~\cite{Wang:2014bba}. All CE$\nu$NS-based 
determinations shown in the figure inset are mutually 
consistent and in good agreement with the SM 
prediction, painting a coherent picture of the 
electroweak sector across a wide range of momentum transfers.

\subsection{Joint fit of \boldmath$R_n(\mathrm{Ge})$ and 
\boldmath$\sin^2\vartheta_W$}
\label{sec:2D}

\begin{figure}[t]
    \includegraphics[width=\linewidth]{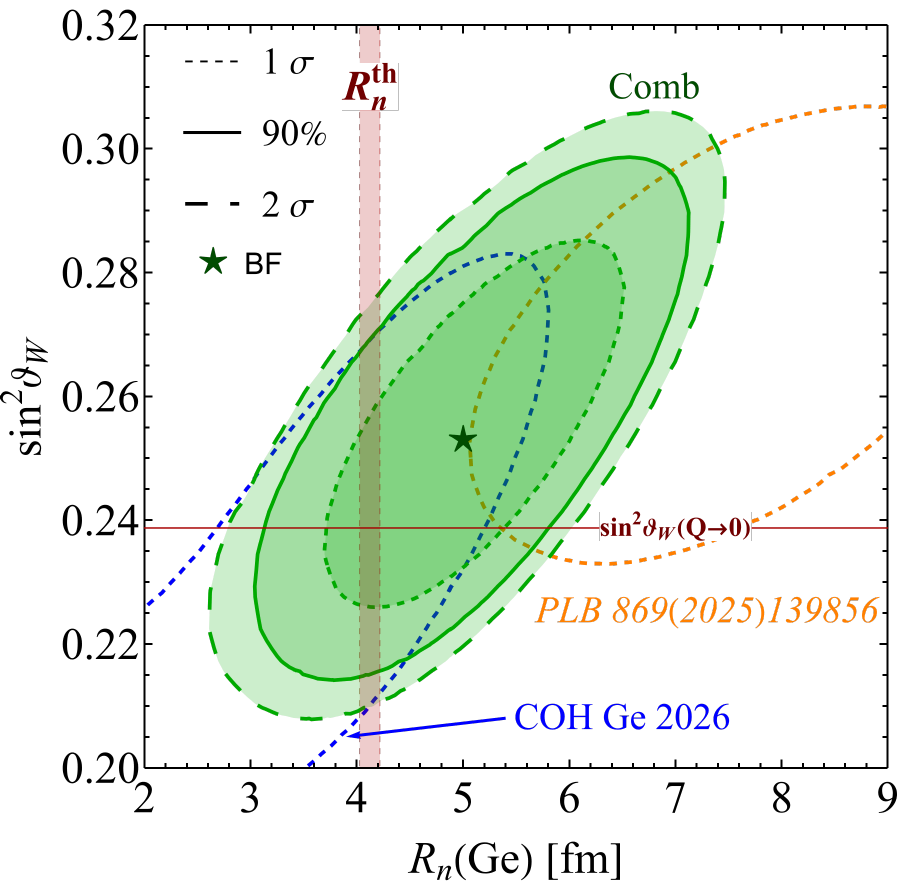}
    \caption{Constraints obtained by fitting simultaneously the weak mixing angle and the average rms Ge neutron radius on COHERENT Ge 2026 data (blue contour), the combination~\cite{AtzoriCorona:2025xgj} of \texttt{Reactors}, $\nu-e^-$ scattering data and the COHERENT germanium 2025  result (orange 
    contour) and the full combination (green regions). Contours are at 1$\sigma$ (dotted), 
    90\% (solid) and 2$\sigma$ (dashed) CL. The theoretical predictions are shown by the red bands. }
    \label{fig:2D_Rn_sin2w}
\end{figure}

We next consider the simultaneous determination of $R_n(\mathrm{Ge})$ 
and $\sin^2\vartheta_W$, treating both as free parameters in the fit. 
The resulting confidence regions in the ($R_n(\mathrm{Ge})$, $\sin^2\vartheta_W$) 
plane are shown in Fig.~\ref{fig:2D_Rn_sin2w} at $1\sigma$, 90\% 
and $2\sigma$ CL, for the analysis of COHERENT Ge~2026 alone and for the combination with the previous 
COHERENT germanium dataset, \texttt{Reactors}~\cite{AtzoriCorona:2025xgj} and $\nu-e^-$ scattering data from TEXONO~\cite{TEXONO:2009knm}, LSND~\cite{LSND:2001akn}, LAMPF~\cite{Allen:1992qe}, LZ~\cite{LZ:2022lsv}, PandaX~\cite{PandaX:2022ood} and XENONnT~\cite{XENON:2022ltv}. For reference, the SM value 
$\sin^2\vartheta_W(Q \to 0)$ is indicated by the horizontal red line, and the theoretical 
prediction for $R_n(\mathrm{Ge})$ 
is shown as a vertical shaded band.

The two-dimensional fit confirms the picture emerging from the 
one-dimensional analysis: the new COHERENT germanium data 
alone already yield a competitive determination of both parameters. The combination 
with \texttt{Reactor} data, which provides a tight, largely 
$R_n$-independent constraint on $\sin^2\vartheta_W$, 
breaks the degeneracy between the two parameters that 
is intrinsic to the accelerator-only fit, resulting in 
a significant reduction of the allowed region. The 
combined two-dimensional best-fit values, at 1 $\sigma$, are
\begin{align}
    \textrm{2D Comb:}\,\sin^2\vartheta_W &= 0.253^{+0.022}_{-0.018}\,,\\
    \textrm{2D Comb:}\, R_n(\mathrm{Ge}) &= 5.0^{+1.0}_{-0.8}\;\mathrm{fm}\,.
\end{align}
Driven by the positive correlation between $R_n(\mathrm{Ge})$ and $\sin^2\vartheta_W$ inherent to the COHERENT data, the extracted neutron rms radius takes on a relatively large value. Nevertheless, it remains compatible with HF+BCS calculations within the current uncertainties~\cite{Hoferichter:2020osn,AtzoriCorona:2025xgj}. The combined result supersedes the previous 
determination of Ref.~\cite{AtzoriCorona:2025xgj}, yielding 
a markedly reduced uncertainty driven by the improved 
statistical precision of the 2026 dataset.

\subsection{Neutrino charge radii}
\label{sec:CR}

\begin{figure}[t]
\includegraphics[width=\linewidth]{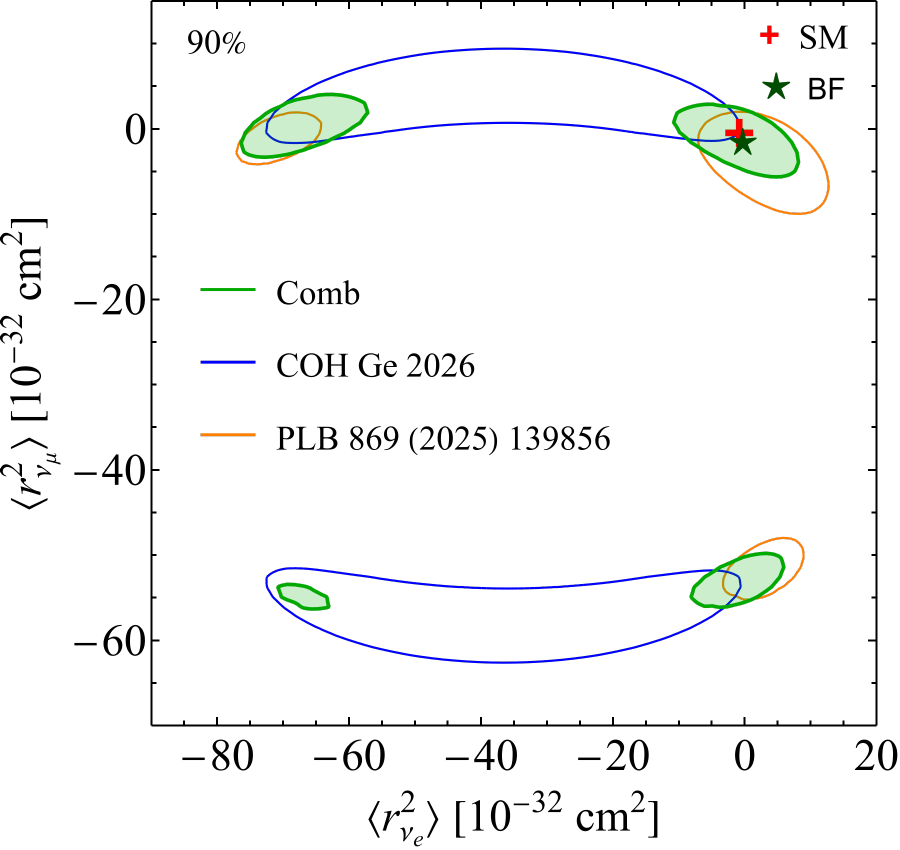}
    \caption{Allowed regions at 90\% CL on the electronic and muonic neutrino charge radii, from the analysis of COHERENT 2026 germanium (blue contour), the combination~\cite{AtzoriCorona:2025xgj} of COHERENT CsI, Ar, and 2025 Ge as well as \texttt{Reactor} CE$\nu$NS data (orange contour), together with their combined analysis (green shaded area). The red cross indicates the SM values reported in Eqs.~(\ref{eq:cr-e}) and (\ref{eq:cr-mu}), while the green star represents the best-fit of the combined analysis.}
    \label{fig:Rcharge}
\end{figure}

We extract the neutrino charge radii 
$\langle r^2_{\nu_e} \rangle$ and 
$\langle r^2_{\nu_\mu} \rangle$ by treating 
them as free parameters in the fit and considering a momentum-dependence in the neutrino charge radii radiative corrections as detailed in Ref.~\cite{AtzoriCorona:2024rtv}. The 
corresponding 90\% confidence regions in 
the $(\langle r^2_{\nu_e} \rangle, 
\langle r^2_{\nu_\mu} \rangle)$ plane are shown 
in Fig.~\ref{fig:Rcharge} for COHERENT Ge~2026, together with the previous CE$\nu$NS result reported in Ref.~\cite{AtzoriCorona:2025xgj}, and their combination. 
For each neutrino flavor, there 
exist two values of the charge radius that produce the same 
observed rate: one close to the SM prediction and one 
substantially shifted. As expected, 
this fourfold degeneracy structure characteristic of neutrino charge 
radii measurements~\cite{Cadeddu:2018dux} also appears here, 
with two/four allowed regions symmetric 
under swap of both charge radii. The 
combination selects the physically preferred 
solution in agreement with the SM expectation~\cite{Bernabeu:2000hf,Bernabeu:2002nw}
\begin{align}
\langle{r}_{\nu_{e}}^2\rangle_{\text{SM}} &\simeq -0.83 \times 10^{-32} \, \text{cm}^2, \label{eq:cr-e} \\
\langle{r}_{\nu_{\mu}}^2\rangle_{\text{SM}} &\simeq -0.48 \times 10^{-32} \, \text{cm}^2. \label{eq:cr-mu}
\end{align}
Here, we report the combined best-fit values in the region around the SM, as the degenerate regions are excluded by a global fit of neutrino data~\cite{AtzoriCorona:2025xwr}, namely
\begin{align}
    \langle r^2_{\nu_e} \rangle \!&= 
    -0.3^{+3.7}_{-4.7}(1\sigma),^{+6.4}_{-7.8}(90\%)\times 10^{-32}\;\mathrm{cm}^2,\\ 
    \langle r^2_{\nu_\mu} \rangle \!&= 
    -1.6^{+3.1}_{-1.7}(1\sigma),^{+4.6}_{-3.1}(90\%)
    \times 10^{-32}\;\mathrm{cm}^2.
\end{align}
\section{Implications for Quenching Factor Models}\label{sec:QFModels}

The quenching factor is a fundamental property of detector materials, as it encapsulates the response to nuclear recoils relative to that of electron recoils depositing the same amount of energy. Consequently, it has been extensively studied by various experiments over the years~\cite{PhysRevA.11.1347,PhysRevC.4.125,CoGeNT:2012sne,SuperCDMS:2022nlc,Bonhomme:2022lcz,Collar:2021fcl}. Nevertheless, results still show non-negligible discrepancies, particularly at low energies (see e.g. Fig.~1 of Ref.~\cite{SuperCDMS:2022nlc}), despite an overall agreement with the Lindhard theory. In this section we use the 
new COHERENT germanium data to perform a direct test of the 
quenching factor model adopted in our baseline analysis. In particular, as experiments continue to lower their detection thresholds, achieving a precise understanding of the quenching factor in the low-energy regime becomes increasingly important for the interpretation of the data~\cite{Li:2025pfw,Liao:2021yog,AtzoriCorona:2023ais,AristizabalSierra:2022axl}. To this end, here we consider the extension of the Lindhard theory proposed in Ref.~\cite{Sorensen:2014sla}, which parametrizes possible additional microscopic effects contributing to the ionization yield through an extra term $q$, namely
\begin{equation}\label{eq:MLT}
    f_Q=\frac{k\;g(\epsilon)}{1+k\;g(\epsilon)}-\frac{q}{\epsilon}\, , 
\end{equation}
where \mbox{$k \simeq 0.133\;Z^{2/3}\;A^{-1/2}$} (which equals $k\simeq0.157$ for Ge) determines the scaling factor in the energy loss equation, while $g(\epsilon)\simeq3\epsilon^{0.15}+0.7\epsilon^{0.6}+\epsilon$ with $\epsilon\simeq11.5\;Z^{-7/3}T_{\rm nr}$ and $T_{\rm nr}$ is expressed in units of $\rm keV$. The standard Lindhard theory is retrieved by employing the above definitions for $k,\, g(\epsilon)$ and $\epsilon$ and fixing $q = 0$, as discussed in Sec.~\ref{sec:IntroCohGe}.
\begin{figure}[t!]
    \centering
    \includegraphics[width=0.98\linewidth]{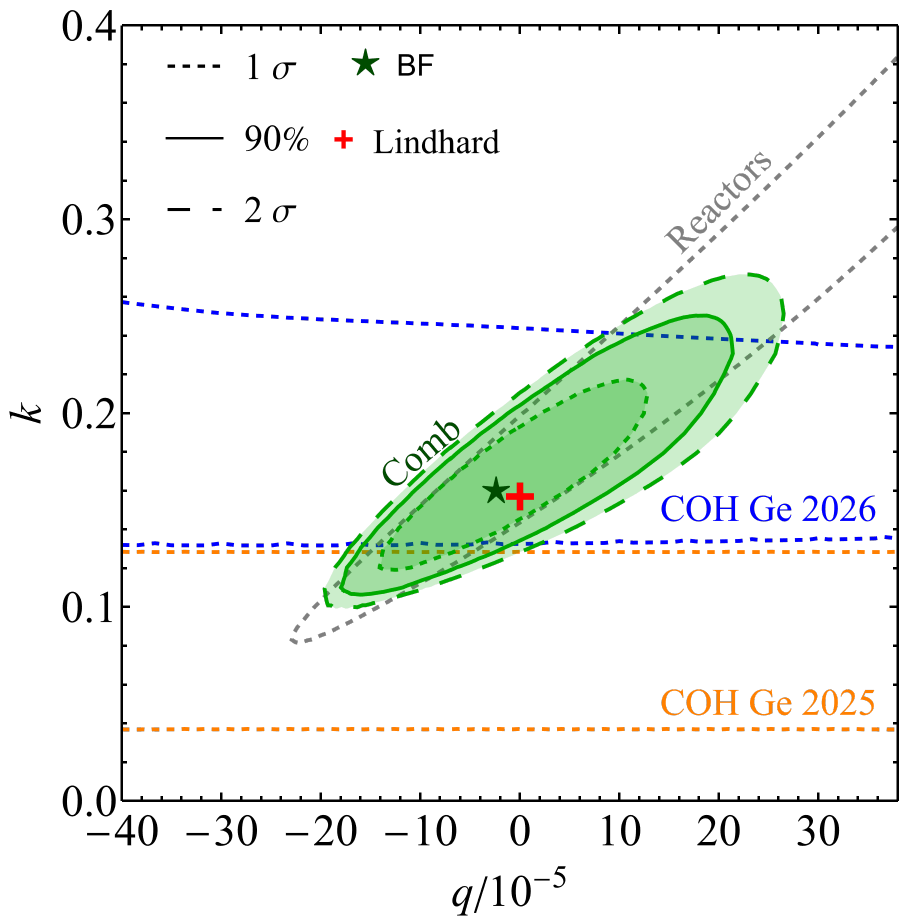}
    \caption{Constraints on the parameters $(k,q)$ of the extended Lindhard quenching factor model obtained from CE$\nu$NS data on germanium. Allowed regions are shown at $1\sigma$ (dotted), $90\%$ (solid), and $2\sigma$ (dashed) CL for \texttt{Reactor} data (grey), COHERENT Ge 2025 (orange) and 2026 (blue), and their combination (green shaded area). The star indicates the best-fit, while the red cross marks the standard Lindhard prediction, \textit{i.e.} for $k=0.157$ and $q=0$.}
    \label{fig:ModLind2D}
\end{figure}
To test the quenching model, we allow both $k$ and $q$ to vary 
freely in the fit while fixing the CE$\nu$NS cross section to 
its SM prediction. The two parameters affect the ionization 
yield in distinct and complementary ways: $k$ controls the 
overall normalisation of $f_Q$, producing a roughly uniform 
rescaling of the ionization yield, whereas $q$ governs its 
low-energy behaviour, making low-threshold experiments, such as those at reactors, 
particularly sensitive to its value~\cite{Liao:2022hno}. In this procedure, the quenching factor systematic uncertainty is removed from the experimental systematics, as the quenching factor parameters $k$ and $q$ are directly left free to vary in the fit.
The result is depicted in Fig.~\ref{fig:ModLind2D}, which shows a $90\%$ CL agreement between the expected Lindhard parameters and the result of the combined fit. Specifically, the combined analysis gives
\begin{align}
    k \!&= 
    0.160^{+0.036}_{-0.029}(1\sigma),^{+0.065}_{-0.044}(90\%),^{+0.082}_{-0.052}(2\sigma)\, ,\\ 
    q&= 
    -2^{+10}_{-8}(1\sigma),^{+17}_{-13}(90\%),^{+22}_{-15}(2\sigma)
    \times 10^{-5}\;.
\end{align}
The result of this fit, obtained solely using neutrino CE$\nu$NS data, can be qualitatively compared with the result of the neutron calibration campaign conducted by the CONUS+ collaboration, namely $k=0.162\pm0.004$~\cite{Bonhomme:2022lcz}, and represents an independent determination of the germanium quenching factor.
It is worth noting that this model was already studied in Ref.~\cite{Liao:2022hno} in the context of CE$\nu$NS experiments, with the aim of explaining the unexpectedly high ionization yield observed during the calibration of the Dresden II detector~\cite{Colaresi:2022obx}. We analyzed the Dresden II data as detailed in Ref.~\cite{AtzoriCorona:2022qrf}, exploring a wider parameter space compared to Ref.~\cite{Liao:2022hno}. However, we found that the $1\sigma$ contour lies outside the bounds of Fig.~\ref{fig:ModLind2D}, clearly showing tension with the other data; therefore, we do not include this dataset in the combined analysis.\footnote{For a detailed discussion of the compatibility of Dresden II data with the other reactor experiments, please refer to Ref.~\cite{Li:2025pfw}.}
\begin{figure}[t!]
    \centering
    \includegraphics[width=0.98\linewidth]{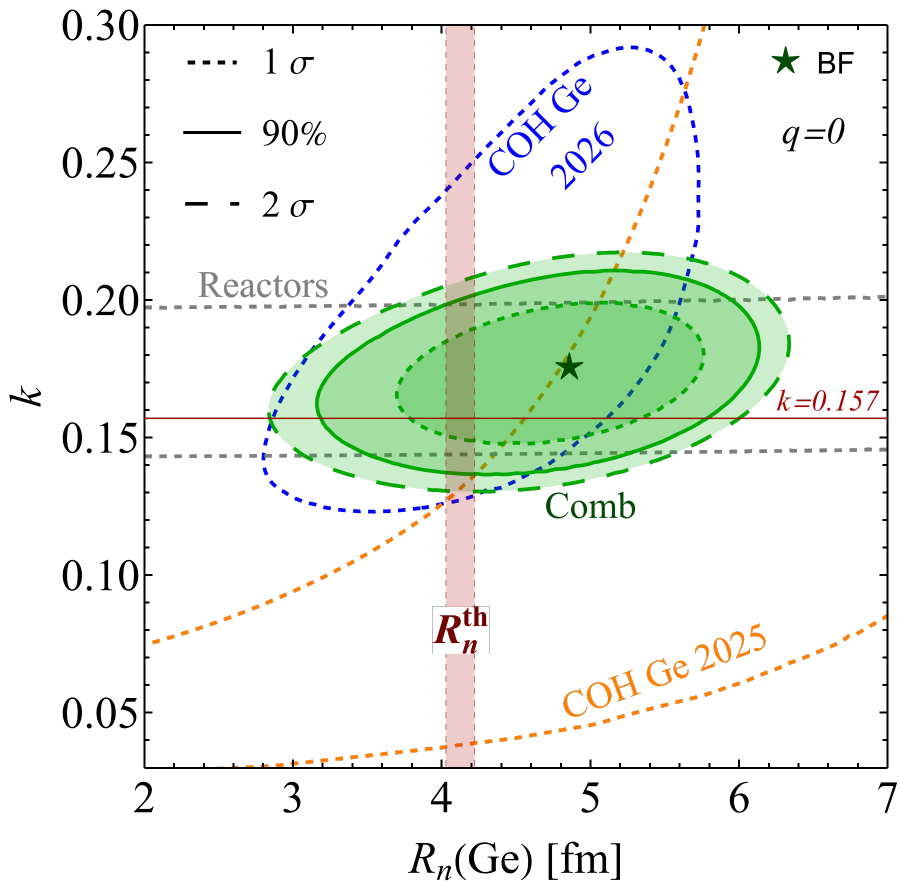}
    \caption{Correlation between the Lindhard parameter $k$ and the neutron radius $R_n(\mathrm{Ge})$, fixing $q=0$. The blue contours show the allowed regions from the COHERENT Ge~2026 dataset, the orange curves correspond to the COHERENT Ge~2025 data, while the gray band corresponds to \texttt{Reactors} data. The green shaded regions represent the combined analysis, and the star indicates the global best-fit point. Contours are shown at $1\sigma$ (dotted), $90\%$ CL (solid), and $2\sigma$ (dashed). The horizontal red line marks the standard Lindhard value $k=0.157$, while the vertical shaded band indicates the theoretical predictions for the germanium neutron rms radius.}
    \label{fig:kLindRn2D}
\end{figure}
\\
While phenomenological analyses typically assume that the standard Lindhard parameterization provides an accurate description of $f_Q$, the question of its universality remains open. In particular, small detector-dependent effects, arising, for instance, from the specific crystal properties, impurity content, operating temperature, or detector-specific conditions, may be present and could potentially bias the extraction of physical parameters. For this reason, we investigate the correlation between the standard Lindhard parameter $k$ and the average neutron rms radius of germanium when fixing $q=0$, in order to assess whether deviations from the reference $k$ value can have a non-negligible impact on the determination of $R_n (\mathrm{Ge})$. The results of our fit are shown in Fig.~\ref{fig:kLindRn2D}, which clearly reveals a correlation in the $k-R_n(\mathrm{Ge})$ plane. In particular, a smaller value of $k$, and hence a smaller quenching factor in the relevant recoil-energy range, leads to a suppression of the expected event rate, which can be compensated for by a smaller neutron radius.
The combination of the data from \texttt{Reactors} and COHERENT yields a value of $k$ consistent with the Lindhard prediction,
\begin{align}
    &k=0.175^{+0.014}_{-0.019}(1\sigma)\,,^{+0.026}_{-0.030}(90\%)\,,^{+0.032}_{-0.037}(2\sigma)\, ,\\
   &R_n(\mathrm{Ge})= 4.9^{+0.6}_{-0.8}(1\sigma),^{+1.0}_{-1.3}(90\%),^{+1.2}_{-1.6}(2\sigma)\;\mathrm{fm}\,.
\end{align}
A direct comparison with the nominal result reported in Eq.~\eqref{eq:Rn_comb} 
($R_n(\text{Ge}) = 4.63_{-0.69}^{+0.63}$~fm)  
highlights the sensitivity of the extracted radius to the treatment of the quenching factor. Indeed, profiling over $k$ leads to a slight upward shift in the central value and a broadening of the uncertainties. This demonstrates that, to achieve high-precision extractions in future analyses, an accurate determination of the quenching factor and a significant reduction of its associated uncertainties will be crucial.

\section{Implications for Physics beyond the Standard Model}

\begin{figure}[tb]
    \centering
    \includegraphics[width=\linewidth]{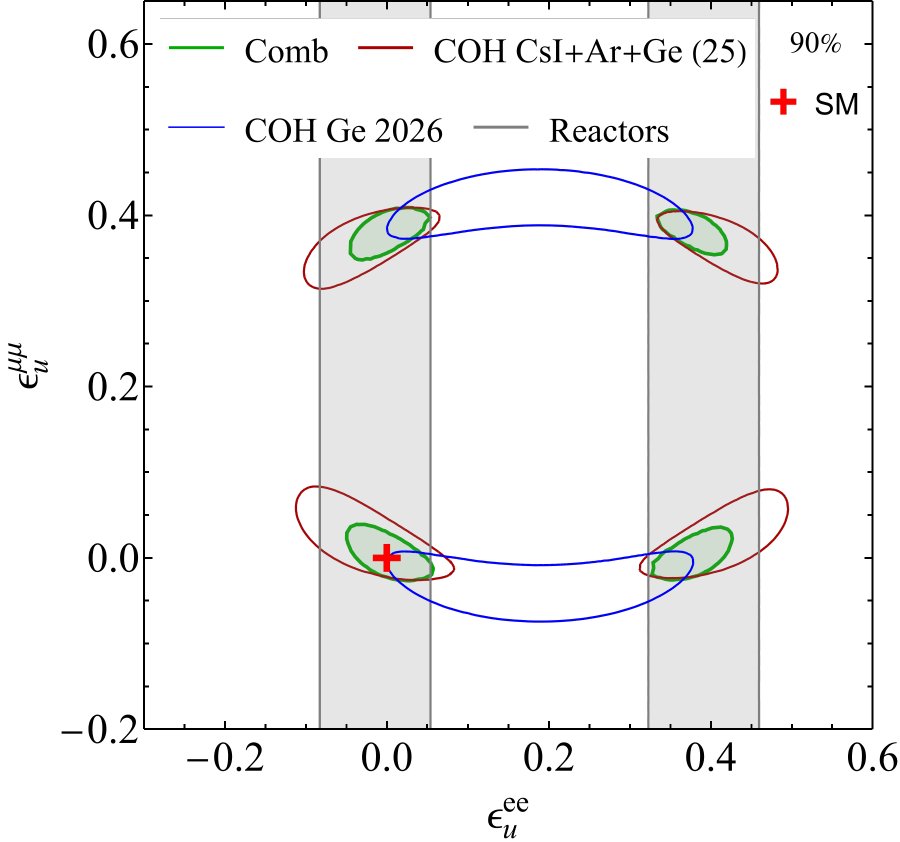}
    \caption{Allowed 90\% CL regions in the $(\epsilon^{ee}_u, \epsilon^{\mu\mu}_u)$ parameter space. The blue contour represents the constraints from the new COHERENT Ge~2026 dataset. The dark red contour shows the combination involving COHERENT CsI, Ar, and the 2025 Ge data, while the grey vertical bands denote the constraints from \texttt{Reactor} experiments, which are sensitive only to the $\nu_e$ flavor. The green filled contours show the full combination, also consistent with the SM expectation (indicated by the red cross).}
    \label{fig:NSI}
\end{figure}

We interpret the data in terms of non-standard neutrino interactions mediated by heavy particles~\cite{Coloma:2017ncl,Coloma:2023ixt,Suliga:2020jfa}, parameterised through generic vector couplings $\epsilon^{\alpha\beta}_q$ affecting the scattering of an $\alpha$-flavor neutrino off a $q$-type quark. For simplicity, we consider a flavor-diagonal scenario ($\alpha=\beta$) involving only the neutrino-up-quark scattering by performing a two-dimensional fit in the $(\epsilon^{ee}_u, \epsilon^{\mu\mu}_u)$ plane. The resulting 90\% confidence regions are shown in Fig.~\ref{fig:NSI} for COHERENT Ge~2026 alone, COHERENT CsI+Ar+Ge~2025 and \texttt{Reactors}, and the full combination.
The COHERENT Ge~2026 contour exhibits the characteristic two/four-lobe structure that arises from the quadratic dependence of the CE$\nu$NS cross section on the effective neutrino-quark couplings: a flip in both $\epsilon^{ee}_u$ and $\epsilon^{\mu\mu}_u$ can leave the total rate approximately invariant, generating two pairs of degenerate solutions symmetric under $\epsilon^{ee}_u \to \bar{\epsilon} - \epsilon^{ee}_u$, where $\bar{\epsilon}$ is determined by the SM weak charge. The \texttt{Reactors} constraint, being sensitive exclusively to the $\nu_e$ flavor, appears as two vertical bands that restrict the allowed values of $\epsilon^{ee}_u$ independently of $\epsilon^{\mu\mu}_u$. The combination of all datasets retains four compact regions, one of which centred on the SM point $(\epsilon^{ee}_u, \epsilon^{\mu\mu}_u) = (0, 0)$. While the constraints derived using solely the COHERENT Ge 2026 data are fully consistent with the official bounds reported by the collaboration~\cite{COHERENT:2026yje}, our global combination with all other available CE$\nu$NS datasets significantly tightens these limits.

\begin{figure}[t!]
    \centering
    \includegraphics[width=\linewidth]{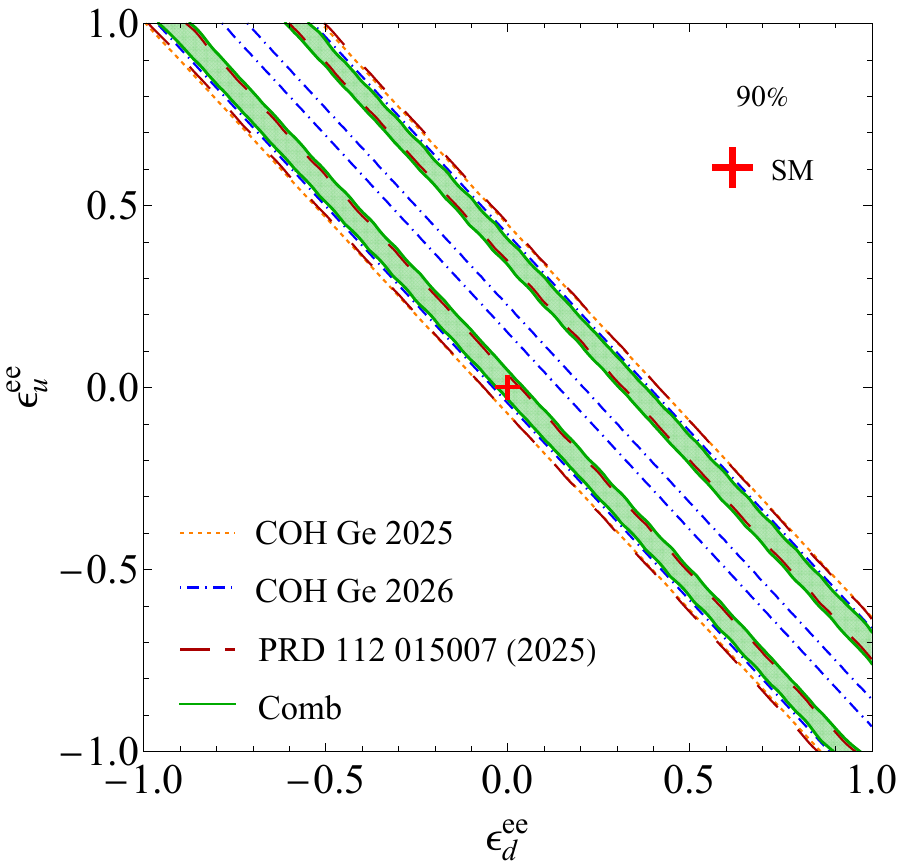}
    \caption{Allowed 90\% CL regions in the $(\epsilon_{d}^{ee}, \epsilon_{u}^{ee})$ parameter space for vector NSI couplings. The plot illustrates the constraints from the previous COHERENT Ge~2025 data (orange dotted lines), the newly updated COHERENT Ge~2026 data (blue dot-dashed lines), and a previous phenomenological fit~\cite{AtzoriCorona:2025ygn} including TEXONO, CONUS+, and COHERENT CsI+Ar (red dashed lines). The solid green shaded region corresponds to the updated global combination. 
    The SM prediction (red cross) lies perfectly within the newly combined allowed region.}
    \label{fig:NSIeued}
\end{figure}

Following the same strategy of Ref.~\cite{AtzoriCorona:2025ygn}, in addition to flavor-diagonal NSI couplings to a specific quark, we also investigate the interplay between NSI couplings to up and down quarks for a given neutrino flavor. Focusing on the electron-neutrino sector, we perform a two-dimensional fit in the $(\epsilon_{d}^{ee}, \epsilon_{u}^{ee})$ plane. The resulting 90\% CL constraints are illustrated in Fig.~\ref{fig:NSIeued}. Because the CE$\nu$NS cross section probes a coherent superposition of neutrino-nucleon scatterings, it is essentially sensitive to a linear combination of the up and down quark couplings weighted by the proton and neutron numbers of the target nucleus. This physical dependence manifests as a strong anti-correlation in the parameter space, generating the observed diagonal bands whose slope is strictly dictated by the $Z/N$ ratio of the target.

As shown in the figure, the latest COHERENT Ge~2026 dataset significantly narrows the allowed parameter space compared to the older COHERENT Ge~2025 measurement, demonstrating the remarkable impact of the increased exposure and lowered analysis threshold. The inclusion of all available datasets~\cite{AtzoriCorona:2025ygn} (TEXONO, CONUS+, and COHERENT CsI+Ar) into a global combined fit provides the most stringent constraints to date along this degenerate direction. Once again, the SM prediction $(\epsilon_{d}^{ee}, \epsilon_{u}^{ee}) = (0,0)$ is thoroughly embedded within the combined 90\% CL allowed region, further reinforcing the robustness of the SM against the presence of non-standard vector interactions in the low-energy regime.

\section{Conclusions}
\label{sec:conclusions}

We have presented the first comprehensive phenomenological analysis 
of the new COHERENT germanium CE$\nu$NS dataset, 
which represents the most precise measurement of this process to date. 
With roughly three times the neutrino exposure of the earlier result, 
a lowered analysis threshold, and improved background rejection, the 
new data cross the threshold from a statistics-dominated to a 
systematics-dominated regime, opening new opportunities for precision 
CE$\nu$NS phenomenology.

Our simultaneous fit to the energy and timing distributions yields $127^{+10}_{-9}$ CE$\nu$NS events, in good 
agreement with both the COHERENT collaboration result and our theoretical 
prediction. 

The combined flavor-separated normalization analysis confirms 
full consistency with the SM within $1\sigma$ for both neutrino 
components, resolving the mild tension observed in the analysis 
of the first germanium dataset~\cite{AtzoriCorona:2025xgj}. We note that when considering the COHERENT Ge 2026 dataset in isolation, the muonic channel exhibits an intriguing excess that warrants further investigation.

From this improved dataset we have extracted updated determinations 
of several SM and nuclear physics parameters. The neutron rms radius 
of germanium is measured as $R_n(\mathrm{Ge}) = 4.11^{+0.75}_{-0.87}$~fm
from COHERENT Ge~2026 alone, while the global CE$\nu$NS combination 
yields $\sin^2\vartheta_W = 0.235^{+0.015}_{-0.012}$, 
the most precise low-energy determination of the weak mixing angle 
from CE$\nu$NS to date.
In addition, we tested the quenching factor in germanium by considering extensions beyond the standard Lindhard framework, constraining both the Lindhard parameter $k$ and an additional parameter $q$ accounting for deviations at low recoil energies. We showed that biases in the quenching factor parameterization can significantly impact the extraction of nuclear physics parameters. In particular, variations in the Lindhard description are correlated with the inferred neutron rms radius, representing a non-negligible source of systematic uncertainty at the current level of experimental precision. Dedicated, experiment-specific quenching factor calibration campaigns would therefore be highly valuable to reduce these systematics and mitigate potential biases in the determination of fundamental parameters.
\\ Constraints on neutrino charge radii and 
on non-standard neutrino interactions in both the 
$(\epsilon^{ee}_u, \epsilon^{\mu\mu}_u)$ and 
$(\epsilon_{d}^{ee}, \epsilon_{u}^{ee})$ planes have also been 
significantly improved with respect to previous analyses, with 
the SM prediction consistently falling within all combined 
allowed regions.

These results demonstrate that CE$\nu$NS has entered a precision 
era. Future improvements in the SNS neutrino flux normalization will further reduce the dominant systematic uncertainty, making 
CE$\nu$NS an increasingly powerful probe of the electroweak 
sector, nuclear structure, and physics beyond the Standard Model.


\bibliography{ref}

\end{document}